\documentclass[a4paper,11pt]{article}
\usepackage{pos}

\title{Scaling properties of
elastic pp
cross-section}

\author*[a]{Micha{\l} Prasza{\l}owicz}
\author[b]{Cristian Baldenegro}
\author[c]{\\ Christophe Royon}
\author[d]{Anna M. Sta{\'s}to}

\affiliation[a]{Institute of Theoretical Physics, Faculty of Physics, Astronomy and Applied Computer Science, \\
Jagiellonian University, S. {\L}ojasiewicza 11,
30-348 Krak{\'o}w, Poland}
\affiliation[b]{Dipartimento di Fisica, Sapienza Universit{\' a} di Roma, \\ Piazzale Aldo Moro, 2, 00185 Rome, Italy}
\affiliation[c]{Department of Physics and Astronomy, \\
The University of Kansas, Lawrence, KS 66045, USA}
\affiliation[d]{Department of Physics, Penn State University, \\University Park, PA 16802, USA \\~~}

\emailAdd{michal.praszalowicz@uj.edu.pl}
\emailAdd{c.baldenegro@cern.ch}
\emailAdd{Christophe.Royon@cern.ch}
\emailAdd{ams52@psu.edu}

\abstract{We show that the elastic differential $pp$ cross-section has a unique universal property that the ratio of bump-to-dip position 
is constant from the energies of the ISR to the LHC. We explore this property to compare Geometric Scaling present 
at the ISR with the recently proposed scaling law at the LHC. We argue that at the LHC, within present experimental uncertainties, 
there is fact a family of scaling laws. 
}

\FullConference{31st International Workshop on Deep Inelastic Scattering (DIS2024)\\
 8–12 April 2024\\
Grenoble, France\\}

\begin{document}
\maketitle

\section{Introduction}

In this report  we discuss scaling properties of the differential elastic $pp$ cross-section
at low and high energies following Refs.~\cite{Baldenegro:2022xrj,Baldenegro:2024vgg}.
Differential $pp$ cross-sections have very characteristic structure. One observes a rapid decrease for  small $|t|$, then a minimum at $t_{\rm d}$, called a {\it dip},
followed by a broad maximum at $t_{\rm b}$, dubbed as a {\it bump}.
It turns out that the bump-to-dip cross-section ratio
\begin{equation}
\mathcal{R}_{\mathrm{bd}}(s)=\frac{  d\sigma_{\mathrm{el}} /d|t|_{\rm{b}}}
{ d\sigma_{\mathrm{el}} /d|t|_{\mathrm{d}}}\, ,%
\label{eq:Rbd}%
\end{equation}
which seems to saturate at LHC energies
 at a value of approximately 1.8,  is rather strongly energy
dependent at the ISR (see e.g. Fig.~2 in Ref.~\cite{TOTEM:2020zzr} ).

In Ref.~\cite{Baldenegro:2024vgg} we explored an interesting regularity of $ d\sigma_{\mathrm{el}} /d|t|$, which,
to the best of our knowledge, has not been used in phenomenological studies of the $pp$ elastic
scattering. It turns out that the ratio of  bump-to-dip {\em positions} in $|t|$
\begin{equation}
\mathcal{T}_{\rm bd}(s)=|t_{\rm b}|/|t_{\rm d}|\, ,
\label{eq:Tbd}
\end{equation}
is constant, within the experimental uncertainties, at all energies from the ISR to the LHC and equal approximately 
to 1.355~\cite{Baldenegro:2024vgg}. It is striking and unexpected that this ratio is constant over such large range
of energies. 

In the following, w discuss first Geometric Scaling (GS) at the ISR, and then new scaling laws at the LHC.
For details see Refs.~~\cite{Baldenegro:2022xrj,Baldenegro:2024vgg}.

\section{Geometric Scaling at the ISR}
\label{sec:GS}

Fifty years ago, in 1973-74, Jorge Dias de Deus proposed and developed the idea of GS
in elastic $pp$ scattering \cite{DiasDeDeus:1973lde,Buras:1973km}. It was based on a phenomenological
observation that elastic, inelastic and total cross-sections have (almost) the same energy dependence over the
ISR energy span, see Tab.~\ref{tab:sigmas}.

\renewcommand{\arraystretch}{1.5} \begin{table}[h]
\centering
\begin{tabular}
[c]{|c|c|c|c|c|c|}\hline
& elastic & inelastic & total & $\frac{\mathrm{elastic}}{\mathrm{inelastic}}$
& $\rho$\\\hline
ISR & $W^{0.1142\pm0.0034}$ & $W^{0.1099\pm0.0012}$ & $W^{0.1098\pm0.0012}$ & $W^{0.0043\pm 0.0036}$ & $0.02-0.095$%
\\\hline
LHC & $W^{0.2279\pm0.0228}$ & $W^{0.1465\pm0.0133}$ & $W^{0.1729\pm 0.0163}$ & $W^{0.0814 \pm 0.0264}$ & $0.15-0.10$%
\\\hline
\end{tabular}
\caption{Energy dependence of the integrated cross-sections for the energies
$W=\sqrt{s}$ at the ISR \cite{Amaldi:1979kd}
and at the LHC \cite{Nemes:2019nvj} and the $\rho$ parameter \cite{Amaldi:1979kd,TOTEM:2017asr}.
Fits from \cite{Baldenegro:2024vgg}.}
\label{tab:sigmas}%
\end{table}\renewcommand{\arraystretch}{1.0}

Elastic cross-sections can be parametrized in terms of the opacity $\Omega(s,b)$ and phase $\chi(b,s)$, 
which are both real functions of the impact parameter and  scattering energy \cite{Barone:2002cv,Levin:1998pk}
\begin{equation}
\sigma_{\mathrm{el}}   =
{\displaystyle\int}
d^{2}\boldsymbol{b}\,\left\vert 1-e^{-\Omega(s,b)+i\chi(s,b)}\right\vert
^{2} \, ,
~~~~~~
\sigma_{\mathrm{inel}}   =%
{\displaystyle\int}
d^{2}\boldsymbol{b}\,\left[  1-\left\vert e^{-\Omega(s,b)}\right\vert 
^{2}\right] 
\label{eq:sigmas}
\end{equation}
and
\begin{equation}
\sigma_{\mathrm{tot}}   =2%
{\displaystyle\int}
d^{2}\boldsymbol{b}\,\operatorname{Re}\left[  1-e^{-\Omega(s,b)+i\chi
(s,b)}\right] \, .
\label{eq:sigmatot}%
\end{equation}
Since the real part of the scattering amplitude (related to the $\rho$ parameter) is small, one can safely neglect $\chi$.
GS is a hypothesis that 
$\Omega(s,b)=\Omega\left(  b/R(s)\right) 
$
where $R(s)$ is the interaction radius \cite{DiasDeDeus:1973lde} increasing
with energy. Changing the integration variable from $\boldsymbol{b}%
\rightarrow\boldsymbol{B}=\boldsymbol{b}/R(s)$, one obtains that%
\begin{equation}
\sigma_{\mathrm{inel}}=R^{2}(s)%
{\displaystyle\int}
d^{2}\boldsymbol{B}\,\left[  1-\left\vert e^{-\Omega(B)}\right\vert 
^{2}\right] \, ,
\label{eq:siginel}%
\end{equation}
where the integral in (\ref{eq:siginel}) is an energy independent constant. If
we neglect the phase $\chi(s,b)$, both elastic and total cross-sections should scale the same
way, which means that their ratios should be energy independent. As can be seen from Table~\ref{tab:sigmas}, 
this is indeed the case.

GS of total cross-sections has  important consequences for the
differential cross-section:
\begin{equation}
\frac{d\sigma_{\mathrm{el}}}{d|t|}   \sim\left\vert
{\displaystyle\int\limits_{0}^{\infty}}
db^{2}A_{\mathrm{el}}(b^{2},s)\,J_{0}\left(  b\sqrt{\left\vert t\right\vert
}\right)  \right\vert ^{2} 
=\sigma_{\mathrm{inel}}^{2}(s)\left\vert
{\displaystyle\int\limits_{0}^{\infty}}
dB^{2}A_{\mathrm{el}}(B^{2})\,J_{0}\left(  B\sqrt{\tau}\right)  \right\vert
^{2}
\label{eq:diffx}
\end{equation}
where $A_{\mathrm{el}}(b^{2},s)$ is the elastic scattering amplitude and $J_0$
denotes the Bessel function originating from the Fourier transform. Equation (\ref{eq:diffx})
implies that the scaled cross-section 
\begin{equation}
\frac{1}{\sigma_{\mathrm{inel}}^{2}(s)}\frac{d\sigma_{\mathrm{el}}}%
{d|t|}(s,t)=\Phi(\tau)
\label{eq:BDdDscaling}%
\end{equation}
should be a universal, energy independent function of the scaling variable $\tau$:
\begin{equation}
\tau=R^{2}(s)|t|\times\mathrm{const.}\, =\sigma_{\mathrm{inel}}(s)\,|t|\, .%
\label{eq:tau}%
\end{equation}
GS  at the ISR was confirmed in
Ref.~\cite{Buras:1973km}, except for the dip region~\cite{DiasdeDeus:1977af},
see also \cite{Baldenegro:2024vgg}. 

\section{Scaling at the LHC}

At the LHC elastic, inelastic and total $pp$ cross-sections have different energy dependence,
see Tab.~\ref{tab:sigmas}, and therefore no GS described in Sect.~\ref{sec:GS}
is expected. Nevertheless, the fact that both (\ref{eq:Tbd}) and (\ref{eq:Rbd}) saturate
at the LHC energies, suggests that the transformation
\begin{equation}
t \rightarrow \tau= f(s) t
\end{equation}
should align dips and bumps of different energies, and rescaling
\begin{equation}
\frac{d\sigma_{\mathrm{el}}}{dt}(t) \rightarrow g(s) \frac{d\sigma_{\mathrm{el}}}{dt}(\tau)
\end{equation}
should superimpose the cross-section values, at least in the dip and bump regions. At the ISR
$g\sim 1/f^2$, but at the LHC both functions $f$ and $g$ seem to be independent 
\cite{Baldenegro:2022xrj,Baldenegro:2024vgg}.

\begin{figure}[h]
\centering
\includegraphics[height=6.0cm]{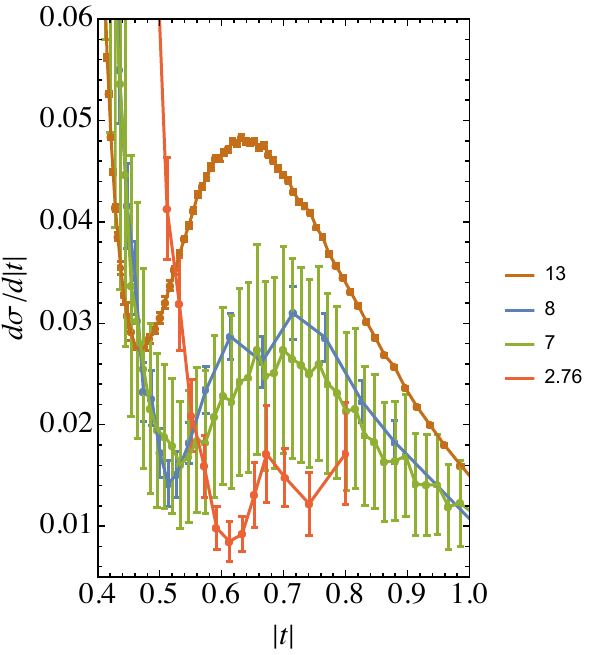}
~\includegraphics[height=6.0cm]{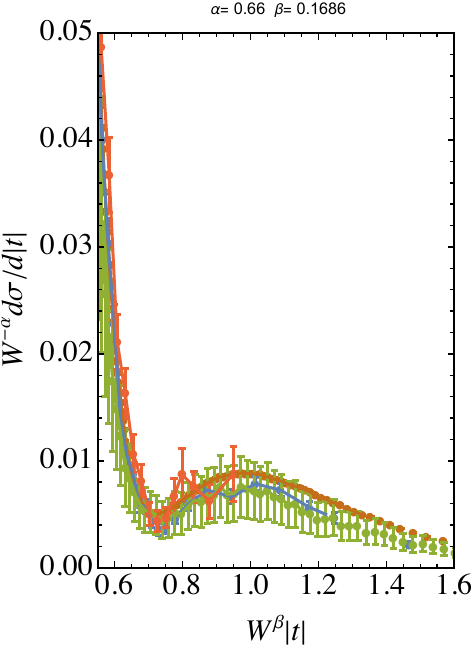}
\caption{Left: elastic $pp$ cross-section
$d\sigma_{\mathrm{el}}/dt$~[mb/GeV$^{2}$] at the LHC energies  in terms of $|t|$~[GeV$^{2}$]  in the dip
and bump region. Right: scaled cross-section in tems
scaling variable $\tau=W^{\protect\beta}\,|t|$.
One can see that the cross-sections at different energies are aligned after scaling.}%
\label{fig:lhc1}%
\end{figure}

In Ref.~\cite{Baldenegro:2024vgg} we have analyzed TOTEM data
\cite{TOTEM:2017asr,TOTEM:2011vxg,TOTEM:2015oop,TOTEM:2017sdy,TOTEM:2018psk} summarized in \cite{Nemes:2019nvj} with the following result:
\begin{equation}
f(s=W^2)=W^\beta,~~\beta=0.1686~~~~~{\rm and}~~~~~g(s=W^2)=W^{-\alpha},~~\alpha\simeq 0.66.
\label{eq:albeta}
\end{equation}
The result is plotted in Fig.~\ref{fig:lhc1} where the cross-section scaling is clearly seen within the experimental uncertainties.

In Ref.~\cite{Baldenegro:2022xrj} a more general scaling variable was proposed
\begin{equation}
\tilde{\tau}=s^{a} t^{b}\label{eq:BRStau}%
\end{equation}
with the result 
and $a \simeq0.065$, $b \simeq0.72$. Aligning dips and bumps alone (rather than the entire cross-sections)
requires the value of $\beta$ given in (\ref{eq:albeta}). This imposes a constraint
\begin{equation}
a-b\, \beta/2 =0,
\label{eq:zeroabbeta}%
\end{equation}
which is roughly satisfied by $a$ and $b$ of Ref.~\cite{Baldenegro:2022xrj}. 
The assumption of energy independence 
of ${\cal T}_{\rm bd}$ (\ref{eq:Tbd}) leads to a family of scaling laws (\ref{eq:BRStau}) 
constrained by (\ref{eq:zeroabbeta})
where the
determination of $a$ (or equivalently of $b$) must follow from the alignment
of the points outside of the immediate vicinity of the dip and bump regions.
The quality of the present LHC data does not allow for a precise determination of $a$ and $b$.

\section*{Acknowledgements}

CB is supported by the European Research Council consolidator grant no. 101002207.
AMS is  supported by the U.S. Department of Energy grant No. DE-SC-0002145 and within the framework of the Saturated Glue (SURGE) 
Topical Theory Collaboration, as well as  in part by National Science Centre in Poland, grant 2019/33/B/ST2/02588.

\end{document}